\documentclass[aps,prl,twocolumn,showpacs,superscriptaddress,floatfix]{revtex4}

\usepackage{graphicx}
\usepackage{epsfig}
\usepackage[english]{babel}

\newcommand{\be}{\begin{equation}}
\newcommand{\bea}{\begin{eqnarray}}
\newcommand{\eea}{\end{eqnarray}}
\newcommand{\ee}{\end{equation}}

\def\one{\ensuremath{\hbox{$\mathrm I$\kern-.6em$\mathrm 1$}}}

\def\qed{\leavevmode\unskip\penalty9999 \hbox{}\nobreak\hfill
     \quad\hbox{\leavevmode  \hbox to.77778em{%
               \hfil\vrule   \vbox to.675em%
               {\hrule width.6em\vfil\hrule}\vrule\hfil}}
     \par\vskip3pt}
\newcommand{\beaa}{\begin{eqnarray*}}
\newcommand{\eeaa}{\end{eqnarray*}}
\newcommand{\bma}{\begin{subequations}}
\newcommand{\ema}{\end{subequations}}

\def\one{{\bf 1}}

\def\noxrightarrow[#1]{\dodoublegroupempty\dodoxrightarrow{#1}}
\def\noxleftarrow [#1]{\dodoublegroupempty\dodoxleftarrow {#1}}
\def\dodoxrightarrow#1#2{\mathrel{{\domthxarr0359\rightarrowfill{#1}{#2}}}}
\def\dodoxleftarrow#1#2{\mathrel{{\domthxarr3095\leftarrowfill{#1}{#2}}}}

\begin{document}

\date{\today}

\title{\bf Quantum circuits for strongly correlated quantum systems}

\author{Frank Verstraete}
\affiliation{Fakult\"at f\"ur Physik, Universit\"at Wien,
Boltzmanngasse 5, A-1090 Wien, Austria.}
\author{J. Ignacio Cirac}
\affiliation{Max-Planck-Institut f\"ur Quantenoptik,
Hans-Kopfermann-Str. 1, D-85748 Garching, Germany.}
\author{Jos\'e I. Latorre}
\affiliation{Dept. ECM, Universitat de Barcelona, Mart\'\i\ i Franqu\`es 1,
08028 Barcelona, Spain.}

\begin{abstract}
In recent years, we have witnessed an explosion of
experimental tools by which quantum systems can be manipulated in
a controlled and coherent way. One of the most important goals now
is to build quantum simulators, which would open up the
possibility of exciting experiments probing various theories in
regimes that are not achievable under normal lab circumstances.
Here we present a novel approach to gain detailed control on the
quantum simulation of strongly correlated quantum many-body
systems by constructing the explicit quantum circuits that
diagonalize their dynamics. We show that the exact quantum
circuits underlying some of the most relevant many-body
Hamiltonians only need a finite amount of local gates. As a
particularly simple instance, the full dynamics of a
one-dimensional Quantum Ising model in a transverse field with
four spins is shown to be reproduced using a quantum circuit of
only six local gates. This opens up the possibility of
experimentally producing strongly correlated states, their time
evolution  at zero time and even thermal superpositions at zero temperature.
Our method also allows to uncover the exact circuits corresponding to
models that exhibit topological order and to stabilizer states.
\end{abstract}
 \maketitle

Recent advances in Quantum Information has led to novel ways of
looking at strongly correlated quantum many-body systems. On the
one hand, a great deal of theoretical work has been made in
identifying the basic structure of entanglement in low-energy
states of many-body Hamiltonians. This has led, for example, to
new interpretations of renormalization group ideas in terms of
variational methods in classes of quantum states with some very
special local structure of entanglement \cite{MPS,PEPS,MERA}, as
well as to new methods to study the low energy properties of
interesting lattice Hamiltonians. On the other hand, new
experimental tools have been developed which should allow us to
simulate certain quantum many--body systems, and thus to gain a
better understanding of their intriguing properties and
potentialities. In particular, the low temperature states
corresponding to the Bose-Hubbard model have been prepared using
atoms in optical lattices \cite{Bloch,Jaksch}, something which has
triggered a lot of attention both in the atomic physics and
condensed matter physics communities.

In this paper we propose to use a quantum computer (or simulator)
in a different way, such that we not only have access to the low
energy states but to the whole spectrum for certain quantum
many--body problems. Moreover, this allows one to prepare any
excited state or thermal state at any temperature, as well as the
dynamical evolution of any state for arbitrary times with an
effort which does not depend on the time, the temperature, or the
degree of excitation. The main idea is to unravel a quantum
circuit that transforms the whole Hamiltonian into one
corresponding to non--interacting particles. As we will show, this
will allow us to achieve the desired goals. Moreover, the new
circuit will be efficient in the sense that the number of gates
only grows polynomially with the number of particles. We will give
some examples where with current systems of 4 or 8 trapped ions it
would be possible to perform a complete simulation of a strongly
interacting Hamiltonian.

We should also mention that what we are doing can be interpreted in
terms of an extension of the renormalization group ideas
\cite{Wilson}. There, one is interested in obtaining a simple
effective Hamiltonian which describes the low energy physics of a
given problem. This is done by a series of transformations which
involve getting rid of high energy modes. In our case, we find a
unitary transformation which takes the whole Hamitonian into a
simple (non-interacting) one, and thus: (i) we do not loose the
physics of the high energy modes in the way; (ii) it can be
implemented experimentally. Of course, our method only works exactly
for the small set of integrable problems, but very similar
approximate transformations can in principle be found for any system
whose effective low-energy physics is well described by
quasi-particles.

Let us start by considering the consequences of identifying the
quantum circuit $\mathcal{U}_{dis}$ that {\em disentangles} a
given Hamiltonian $\mathcal{H}$ acting on $n$ qubits in the
following sense:
 \be
 \mathcal{H}=\mathcal{U}_{dis}\ \tilde{\mathcal{H}}\
 \mathcal{U}_{dis}^\dagger \,
 \ee
where $\tilde{\mathcal{H}}$ is a non--interacting Hamitonian
which, without loss of generality, can be taken as
 \be
 \tilde{\mathcal{H}}=\sum_i \omega_i\sigma_i^z,
 \ee
with $\sigma_i^z$ Pauli operators. We are interested in the
circuits whose size only grows moderately with the number of
qubits. In that case we could: (i) prepare excited eigenstates of
$\mathcal{H}$, just preparing a product state and then applying
$\mathcal{U}_{dis}$; (ii) simulate the time evolution of a
state, just by using
 \be
 e^{-i t \mathcal{H}}=\mathcal{U}_{dis}e^{-i t
 \tilde{\mathcal{H}}}\mathcal{U}_{dis}^\dagger \ .
 \ee
Since $e^{-i t \tilde{\mathcal{H}}}$ can be simulated
using $n$ single--qubit gates, then the whole evolution would
require a fixed number of gates, independent of the time $t$;
(iii) similarly, the same circuit will allow to create the thermal
state $\exp(-\beta \mathcal{H})= \mathcal{U}_{dis}\exp(-\beta
\tilde\mathcal{H})\mathcal{U}_{dis}^\dagger$ explicitly, just by
preparing a product mixed state to start with. This is remarkable,
as, in general, there is no known scheme to create thermal states
using a (zero-temperature) quantum computer.

The goal is thus, to identify the quantum circuits that
disentangle certain kind of Hamiltonians. Here we will consider
three kind of Hamliltonians: (i) the XY model; (ii) Kitaev's on
the honey--comb lattice; (iii) the ones corresponding to stabilizer states.
We will concentrate in detail in the first one, since it gives
rise to quantum phase transitions and critical phases, and thus it
is perhaps the most interesting among them. Towards the end of the
paper we will briefly explain how one can carry out the procedure
with the other two.


The circuit $\mathcal{U}_{dis}$ we shall construct is surprisingly
small. For a system of $n$  spins, the total number of gates in
the circuit scales as $n^2$ and the depth of the circuit grows as
$n\log n$. Then, it seems reasonable to envisage experimental
realizations of the quantum circuit $\mathcal{U}_{dis}$ that will
allow to create {\sl e.g.} the ground state of the Quantum Ising
model for any transverse field starting from a trivial product
state and acting only with a small number of local gates. The very
same circuit would produce excited state, superpositions, time
evolution and even thermal states. This is, thus, a quantum
algorithm that could be run in a quantum computer to exactly
simulate a different quantum system.

The strategy to disentangle the XY Hamiltonian is based in tracing
the well-known transformation which solves the model analytically
\cite{JW,LSM}.
The path to follow is divided in three steps: we first need to
implement the Jordan-Wigner map of spins ($\sigma$) into fermions
($c$), then use the Fourier transform to get fermions in momentum
space ($b$) and, finally, perform a Bogoliubov transformation to
completely diagonalize the system in terms of free fermions ($a$).
As we shall discuss in more detail shortly, the first
transformation is just a relabeling of degrees of freedom which
needs no actual action on the system. The fermions $c$ are just an
economical way of carrying along the degrees of freedom that are
subsequently Fourier transformed. On the other hand, both the
Fourier and Bogoliubov transformations are real actions on the
spin degrees of freedom. Thus, the structure of the unitary
transformation that takes the free theory to the original XY
system corresponds to \be \mathcal{U}_{dis}=\mathcal{U}_{FT} \
\mathcal{U}_{Bog} \ee with \be
\mathcal{H}_{Ising}=\mathcal{H}_1[\sigma] \longleftarrow
\mathcal{H}_2[c] \stackrel{\mathcal{U}_{FT}}{\longleftarrow}
\mathcal{H}_3[b] \stackrel{\mathcal{U}_{Bog}}{\longleftarrow}
\mathcal{H}_4[a]=\tilde \mathcal{H} \ee Some of the pieces of the
$\mathcal{U}_{dis}$ transformation may have a very simple form when
view as an action on the coefficients of the wave function. The
problem is however nontrivial in that the Bogoliubov
transformation changes the vacuum and, hence, the problem is
different than the one of simulating a fermionic computer with a
standard quantum computer \cite{fermioniccomputer}.

Let us detail the construction of $\mathcal{U}_{dis}$ for the
XY Hamiltonian
\bea
\label{XYhamiltonian}
\nonumber
&& \mathcal{H}_{XY}=\sum_{i=1}^n\left(\frac{1+\gamma}{2}\sigma^x_i\sigma^x_{i+1}
+\frac{1-\gamma}{2}\sigma^y_i\sigma^y_{i+1}\right)
+\lambda \sum_{i=1}^n\sigma^z_i
\\ &&
+\frac{1+\gamma}{2}\sigma^y_1\sigma^z_2\dots \sigma^z_{n-1}\sigma^y_n +\frac{1-\gamma}{2}\sigma^x_1\sigma^z_2\dots \sigma^z_{n-1}\sigma^x_n \ .
\eea where $\gamma$ parametrizes the $X$-$Y$ anisotropy and $\lambda$ represents the presence of an external transverse magnetic field. The last
two terms above are related to the correct mapping of periodic boundary conditions between spins to fermionic degrees of freedom and are often
dropped as they are suppressed in the large $n$ limit. These terms can also be substituted with the standard periodic terms
$\sigma^x_n\sigma^x_1$ and $\sigma^y_n\sigma^y_1$ for the even total spin-up sector and the same terms with opposite sign in the odd sector. The
Jordan-Wigner transformation
 $c_i=\left(\prod_{m<i} \sigma^z_m\right)(\sigma^x_i-i\sigma^y_i)/2$
is designed to transform the  spin operators into
fermionic modes. This is indeed implemented by
the strings of spin operators which, furthermore, cancel on the Hamiltonian  leading to
\bea
\nonumber
\mathcal{H}_2[c]&=&\frac{1}{2}\sum_{i=1}^n\left((c_{i+1}^\dagger c_i+c^\dagger_i c_{i+1})\right. \\
&&\left. +\gamma (c_i^\dagger c_{i+1}^\dagger+c_i c_{i+1})\right)+\lambda \sum_{i=1}^n
c_i^\dagger c_i \ ,
 \label{hc}
\eea
where $c_{n+1}=c_1$, setting periodic boundary conditions, and
now $c_i,c^\dagger_i$ are fermionic annihilation and creation operators acting on
the vacuum $|\Omega_c\rangle$ as defined by
\be
 \{c_i,c_j\}=0\qquad
\{c_i,c^\dagger_j\}=\delta_{ij}\qquad c_i|\Omega_c\rangle=0\  . \ee Thus, the Jordan-Wigner transformation takes a state of spin 1/2 particles
\be |\psi\rangle=\sum_{i_1i_2...i_n=0,1} \psi_{i_1i_2...i_n}|i_1,i_2,\dots ,i_n\rangle\ , \ee into a fermionic state \be
|\psi\rangle=\sum_{i_1i_2...i_n=0,1} \psi_{i_1i_2...i_n}(c^\dagger_1)^{i_1} (c^\dagger_2)^{i_2}...(c^\dagger_n)^{i_n}|\Omega_c\rangle \ . \ee
The relevant point to observe is that there is no effect on the coefficients $\psi_{i_1i_2...i_n}$. There are no gates to be implemented on the
register in order to reproduce the full dynamics of the system, provided we retain the fact that any further swapping of degrees of freedom will
carry a minus sign from now on. This is a remarkable simplification in our construction.

The first non-trivial part of the quantum circuit for the
XY Hamiltonian is the one associated to the Fourier transform
which must be performed on the fermionic modes
\be
\label{fourier}
b_k=\frac{1}{\sqrt{n}}\sum_{j=1}^{n} e^{i\frac{2\pi}{n} jk}c_j \quad,\quad k=-\frac{n}{2}+1,\ldots,\frac{n}{2} \ .
\ee
This transformation exploits translational invariance and
takes $H_2[c]$ into a  momentum space Hamiltonian
$H_3[b]$.
We here present the construction of this circuit in terms of
two-body local for the
case where $n= 2^k$, that is,  when a classical fast Fourier transform exists,
though the technique has general applicability.
The quantum circuit that produces the above result can be
constructed in the case of $n=8$ as shown in Fig. 1.
\begin{figure}
\begin{center}
\scalebox{0.4} 
{
\includegraphics*{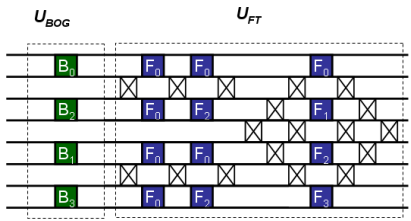}
}
\end{center}
\caption{
Structure of the quantum circuit performing the exact diagonalisation of the
XY Hamiltonian for 8 sites. The circuit follows the structure
of a Bogoliubov transformation followed by a fast Fourier transform.
Three types of gates are involved: type-B (responsible for the
Bogoliubov transformation and depending on
the external magnetic field $\lambda$ and the anisotropy parameter
$\gamma$), type-fSWAP (depicted as crosses and necessary to implement the anticommuting properties of fermions) and type-F
(gates associated to the fast Fourier transform). Some initial gates have been eliminated
since they only amount to some reordering of initial qubits.
}
\label{fig.XX-circuit-n8}
\end{figure}

The circuit contains two types of gates. Every crossing of lines in
the classical fast Fourier transformation corresponds to a fermionic
swap in the quantum case, that we represent with
a crossed box in Fig. \ref{fig.XX-circuit-n8}.
The quantum gate for this fermionic SWAP reads
\be
U_{SWAP}=\left(
    \begin{array}{cccc}
      1 & 0 & 0 & 0 \\
      0 & 0 & 1 & 0 \\
      0 & 1 & 0 & 0 \\
      0 & 0 & 0 & -1 \\
    \end{array}
  \right)
  \label{uSWAP}
\ee
Note the minus sign whenever two ocupied modes enter the fermionic SWAP.
The $\mathcal{U}_{FT}$ circuit also makes use of a second class of
gates. Those implement the change of relative phase associated to the
Fourier transform. The construction of the explicit gates which are needed are
\be
F_k=\left(
    \begin{array}{cccc}
      1 & 0 & 0 & 0 \\
      0 & \frac{1}{\sqrt 2} & \frac{\alpha(k)}{\sqrt 2} & 0 \\
      0 & \frac{1}{\sqrt 2} & -\frac{\alpha(k)}{\sqrt 2} & 0 \\
      0 & 0 & 0 & -\alpha(k) \\
    \end{array}
  \right)\ ,
  \label{uF}
\ee
with $\alpha(k)= \exp(i 2\pi k/n)$.
The fast Fourier classical circuit
is of depth $n \log n$. The quantum circuit needs further
fermionic swaps that makes the total number of gates to grow as
$n^2$. More precisely, the counting of gates in the circuit goes as follows.
For a system of $n=2^k$ spins, the circuit needs $2^{k-1}(2^k-1)$
local gates. Only $k n$ of them are site-dependent interacting gates, whereas
the rest correspond to fermionic SWAPs needed to ensure the fermionic
character of the effective modes handle in the system.
Let us note that the periodic boundary conditions present
in the system have emerged from a set of initial free modes.
It is the action of gates that builds the appropriate boundary property in the system.

Let us note that the way entanglement builds up in the system
is made apparent in the circuit in Fig. \ref{fig.XX-circuit-n8}.
For instance,
when the system is divided in two sets with four contiguous qubits in each
one  all bipartite entanglement is transmited through four f-SWAP gates.
This is the minimum number of gates necessary to generate the known maximum entanglement along
time evolutions. Thus, no circuit with less gates relating both half chains could
provide an exact solution.

The final step to achieve a full
disentanglement of the XY Hamiltonianin corresponds
to a Bogoliubov transformation .
The momentum-dependent
mixture of modes is disentangled using
\bea
\label{BOGmodes}
\nonumber
a_k&=&\cos(\theta_k/2)b_k-i\sin(\theta_k/2)b_{-k}^\dagger\\
\label{BOGangle}
\theta_k&=&\arccos\left(\frac{-\lambda+\cos\left(\frac{2\pi k}{n}\right)}
{\sqrt{\left(\lambda-\cos\left(\frac{2\pi k}{n}\right)\right)^2+
\gamma^2\sin^2\left(\frac{2\pi k}{n}\right)}}\right).
\eea
This transformation preserves the anticommutation relations.
Within the operators $a_k$, the original Hamiltonian can now be expressed as
\bea \mathcal{H}_4[a]&=&\sum_{k=-n/2+1}^{n/2} \omega_k a_k^\dagger a_k\\
\omega_k&=&\sqrt{\left(\lambda-\cos\left(\frac{2\pi k}{n}\right)\right)^2+
\gamma^2\sin^2\left(\frac{2\pi k}{n}\right)}.
\eea
The Hamiltonian is clearly a sum of noninteracting terms
and its spectrum is equivalent to the completely local spin $1/2$ Hamiltonian
$\tilde\mathcal{H}=\sum_i \omega_i\sigma_i^z$.
Let us note that the above Bogoliubov transformation only mixes pairs of modes.
The precise gate that produces such disentanglement
corresponds to
\be
 B_k=\left(
    \begin{array}{cccc}
      \cos\theta_k & 0 & 0 & i \sin\theta_k \\
      0 & 1 & 0 & 0 \\
      0 & 0 & 1 & 0 \\
      i \sin\theta_k & 0 & 0 & \cos\theta_k \\
    \end{array}
  \right)
  \label{fSWAP}
\ee
with $\theta_k$ given by Eq. (\ref{BOGangle}).
This completes the construction of the circuit
that underlies the XY-Hamiltonian.

\begin{figure}
\begin{center}
{
\includegraphics[width=5cm]{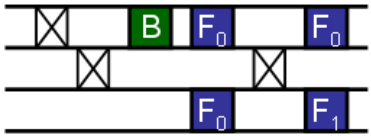} }
\end{center}
\caption{ Complete quantum circuit that reproduces all the dynamical properties of the
Quantum Ising Hamiltonian in a external magnetic field  with four spins.
Note that some initial and final reordering gates could be sparsed
in an experimental realization.} \label{fig.Ising-circuit-n4}
\end{figure}

Let us now discuss the simplest non-trivial experiment that can take
advantage of the circuit we have constructed. We can take
$\gamma=1$, which reduces the system to the quantum Ising chain in a
magnetic transverse field $\lambda$. This theory exhibits a quantum
phase transition for $\lambda=1$ in the $n\to \infty$ limit. Here,
instead, we can consider a system of $n=4$ qubits. Experimentally,
the four qubits should be prepared in the initial $\vert
0000\rangle$ or $|0001\rangle$ state, depending whether $\lambda\leq
1$ or $\lambda>1$ (different valid variants of the circuit we
are presenting change the way the system must be prepared or the
angles appearing in the Bogoliubov transformation). 
Then, the set of gates depicted in the circuit in
Fig.\ref{fig.Ising-circuit-n4} should be operated with a choice of
the parameter $\lambda$ in the only non-trivial Bogoliubov B gate.
To be precise, the angle can be seen to correspond to Eq. \ref{BOGangle}. 
We can further
suppress unnecessary initial and final fermionic swaps since they
just correspond to a relabeling of qubits that can be taken care of
without actual actions on the system.
Actually, only six gates would be needed to recreate the full
dynamics of the Ising model for four qubits!
After running the circuit, the
state of the system would then be the ground state of the Ising
Hamiltonian for that value of the external magnetic field. It would
then be possible to measure {\sl e.g.} the correlator $\langle
\sigma^x_i\sigma^x_j\rangle$, for all $i=1,2,3,4$ and $j\not= i$.
This process could be done over a scan of the $\lambda$ parameter
and scan the magnetization $\langle \sigma^x\rangle$ for any qubit
as well as a measure of three- and four-body correlations functions.

We may finally take a larger view on the underlying structure of the circuits we have presented. The basic idea is that those integrable systems
whose solutions make use of the Jordan-Wigner transformation will have a unitary circuit that disentangles the dynamics with gates of the type
\be V_{ij}=e^{i \alpha (c_i^\dagger c_j + h.c.)} \ , \ W_{ij}=e^{i \alpha (c_i c_j + h.c.)} \ . \ee In our case, the Fourier transform can be
written in terms of $V$-gates, whereas the Bogoliubov transformation needs $W$-gates. These type of gate can further be expressed in terms of
local unitaries because $V_{ij}=V^x_{ij}V^y_{ij}$ with \bea
&&V^x_{ij}=e^{i \alpha \sigma^x_i\sigma^z_{i+1}\dots \sigma^z_{j-1}\sigma^x_j}\\
&&V^y_{ij}=e^{i \alpha \sigma^y_i\sigma^z_{i+1}\dots \sigma^z_{j-1}\sigma^y_j}  \ ,
\eea
and, similarly, $W_{ij}=W_{ij}^x W^y_{ij}$, with
\bea
&&V^x_{ij}=e^{i \alpha \sigma^x_i\sigma^z_{i+1}\dots \sigma^z_{j-1}\sigma^x_j}\\
&&V^x_{ij}=e^{i \alpha \sigma^x_i\sigma^z_{i+1}\dots \sigma^z_{j-1}\sigma^x_j}
\eea
Then, all these gates can be implemented using B-type, F-type and fermionic SWAP gates,
as described previously.

The method we have presented here can be extended to solve other
quantum systems of relevance. Let us sketch two specific cases.
First, we focus on the 2-dimensional Kitaev Hamiltonian on the
honeycomb lattice \cite{Kitaev}. That Hamiltonian is particularly
interesting because its ground state exhibits nontrivial topological
features and can nevertheless be solved exactly using a mapping to
free Majorana fermions. The construction of the quantum circuit
diagonalizing the Hamiltonian can be constructed in the same way as
for the Ising Hamiltonian. The only difference is that, due to the
mapping of one spin 1/2 to two fermions or 4 Majorana fermions,
ancilla's have to be used in the quantum circuit; but these can
again simply be disentangled at the end. A second example of system
whose exact circuit can be obtained corresponds to the case of
stabilizer states \cite{Gottesman}. This class of states is
particularly interesting from the point of view of condensed matter
theory as it encompasses the toric code state and all the so--called
string net states as arising in the context of topological quantum
order \cite{Wen}. The related quantum Hamiltonian is a sum of
commuting terms, each term consisting of a product of local Pauli
operators. Such a Hamiltonian can always be diagonalized by a
quantum circuit only consisting of Clifford operations
\cite{GottesmanKnill}. In principle, those gates can be highly
nonlocal, and in the case of Hamiltonians exhibiting topological
quantum order, one can rigorously prove that the quantum circuit has
a depth that scales linearly in the size of the system \cite{LR}. A
nice measure of the complexity of a particular class of stabilizer
states would be to characterize the minimal depth of the quantum
circuit creating this Hamiltonian.

A more challenging task is to find the quantum circuit
that diagonalizes Hamiltonians that can be solved using the Bethe ansatz. As the
corresponding models are integrable, a quantum circuit is
guaranteed to exist that maps the Hamiltonian to a sum of trivial local terms. Such a
procedure would be very interesting and lead to the
possibility of measuringing correlation functions that are very hard to calculate using the
Bethe ansatz solution.

In conclusion, we have shown that certain relevant Hamiltonians
describing strongly correlated quantum systems can be
exactly diagonalized using a finite-depth quantum circuit. We have
produced the explicit construction of such a circuit that opens up
the possibility of experimental realizations of strongly correlated
systems in controlled devices.

 \vskip 1cm

{\sl Methods.} We here illustrate the technique use to
construct individual quantum gates in the XY.
We consider the example of a fast Fourier transform of four qubits.
The explicit transformations of modes in Eq. (\ref{fourier})
\be
b_k=\left(c_0 + e^{-i2\pi\frac{2k}{4}}c_2\right)+
e^{-i2\pi\frac{k}{4}}\left(c_1 + e^{-i2\pi\frac{2k}{4}}c_3\right)
\ee
where it is made apparent that modes 0 and 2 first mix in the same way as 1 and 3,
and then a subsequent mixing takes place. The first step corresponds to
\bea
\nonumber
&c_0'=c_0+c_2 & c_1'=c_1+c_3\\
&c_2'=c_0+e^{-i \pi}c_2 & c_3'=c_1+e^{-i \pi} c_3
\eea
This the reason why the circuit in Fig. \ref{fig.Ising-circuit-n4}
carries two identical gates in the first part of the Fourier
transform. Similarly, further mixtures will take place after
some fermionic swaps are operated.

To uncover the actual gate needed for the circuit, we consider the wave function made of the modes involve
\bea
\nonumber
\vert\psi\rangle&=&\sum_{i,j=0,1}A'_{ij} \left(c_0^{'\dagger}\right)^i\left(c_2^{'\dagger}\right)^j
\vert 00\rangle\\
& =& \sum_{i,j=0}^1 A_{ij} \left(c_0^\dagger+c_2^\dagger\right)^i\left(c_0^\dagger+e^{i \pi}c_2^\dagger\right)^j
\vert 00\rangle \ .
\eea
Expanding this last equation and working in the basis ${\vert 00\rangle,
\vert 01\rangle,\vert 10\rangle,\vert 11\rangle}$, the coefficients of
the wave function are rearrange by the transformation $A'=U A$ with
\be
\left(
    \begin{array}{cccc}
      1 & 0 & 0 & 0 \\
      0 & \frac{1}{\sqrt{2}} & \frac{1}{\sqrt{2}} & 0 \\
      0 & \frac{1}{\sqrt{2}} & -\frac{1}{\sqrt{2}} & 0 \\
      0 & 0 & 0 & -1 \\
    \end{array}\
  \right) .
\ee

\end{document}